\documentstyle[prl,aps,twocolumn,psfig]{revtex}

\begin{document}

\title{Chaotic hysteresis in an adiabatically oscillating 
double well}
\author{N. Berglund, H. Kunz}
\address{Institut de Physique Th\'eorique,
EPFL, PHB--Ecublens,
CH-1015 Lausanne, Switzerland}
\date{October 3, 1996}

\maketitle

\begin{abstract}
We consider the motion of a damped particle in a potential oscillating 
slowly between a simple and a double well. The system displays hysteresis 
effects which can be of periodic or chaotic type. We explain this 
behaviour by computing an analytic expression of a Poincar\'e map.
\end{abstract}

\pacs{05.45.+b, 64.60.Ht, 64.60.My, 75.60.Nt}


\vspace{-10mm}

Although hysteresis is a quite familiar and ubiquitous phenomenon, no general 
theory achieves to describe its many facets. In condensed matter, hysteresis 
often accompanies a phase transition, which by nature results from the 
cooperative effect of a large number of degrees of freedom. This has 
recently led some workers to analyze it by means of Langevin type \cite{RKP} 
or master equations \cite{LP} with infinitely many degrees of freedom, 
and to propose various scaling laws for the area of the hysteresis loop.
A mean field treatment of this problem \cite{TO} reduces it to an ordinary 
differential equation for the order parameter, with some slowly time 
dependent external parameter like the magnetic field. Noise can be 
incorporated to the problem. Similar equations appear naturally to describe 
mechanical or electrical systems, as well as lasers \cite{ME}. 
Hysteresis effects may appear if the equilibria of the 
dynamical system with static parameter undergo a bifurcation, 
and scaling laws have been found also in this case \cite{JH}.

A paradigmatical example is that of a damped particle in a slowly 
varying potential. If the symmetric potential depends on a parameter 
smoothly interpolating between a simple and a double well, the static 
bifurcation diagramm looks like the inset of Fig.\ 2. Imagine now that 
the parameter is oscillating slowly between these extreme values. 
The particle starting close to the initially stable origin does not 
react immediately to the bifurcation. Rather, it remains for some 
time close to the now unstable origin, before falling into one of 
the newly formed minima, which it follows until it merges again with 
the origin. Thus this {\em bifurcation delay}, which has been observed 
in various physical systems \cite{ME}, and was rigorously analyzed for the 
first time by Neishtadt \cite{Ne}, is responsible for metastability leading 
to hysteresis. 

In this letter we address the following question: for a periodic parameter 
variation, which symmetric potential well will the particle 
fall into during each cycle? In the overdamped case, it always chooses 
the same minimum (Fig.\ 1a). But at low friction, we found that depending 
on the frequency of the parameter variation, the particle may also fall 
alternatively into the left and right equilibrium (Fig.\ 1b), or even go 
from one minimum to the other in a random way (Fig.\ 1c). We observed the 
same phenomenon in an experimental realization of the system, a pendulum 
on a rotating table. In this work, we show that this surprising behaviour 
can be understood by means of a Poincar\'e map, that we compute 
explicitly to lowest order of the parameter variation. We only 
outline the derivation of this fairly complicated, although essentially 
one--dimensional map, postponing rigorous proofs to a further 
publication \cite{BK}.

\begin{figure}
\centerline{\psfig{figure=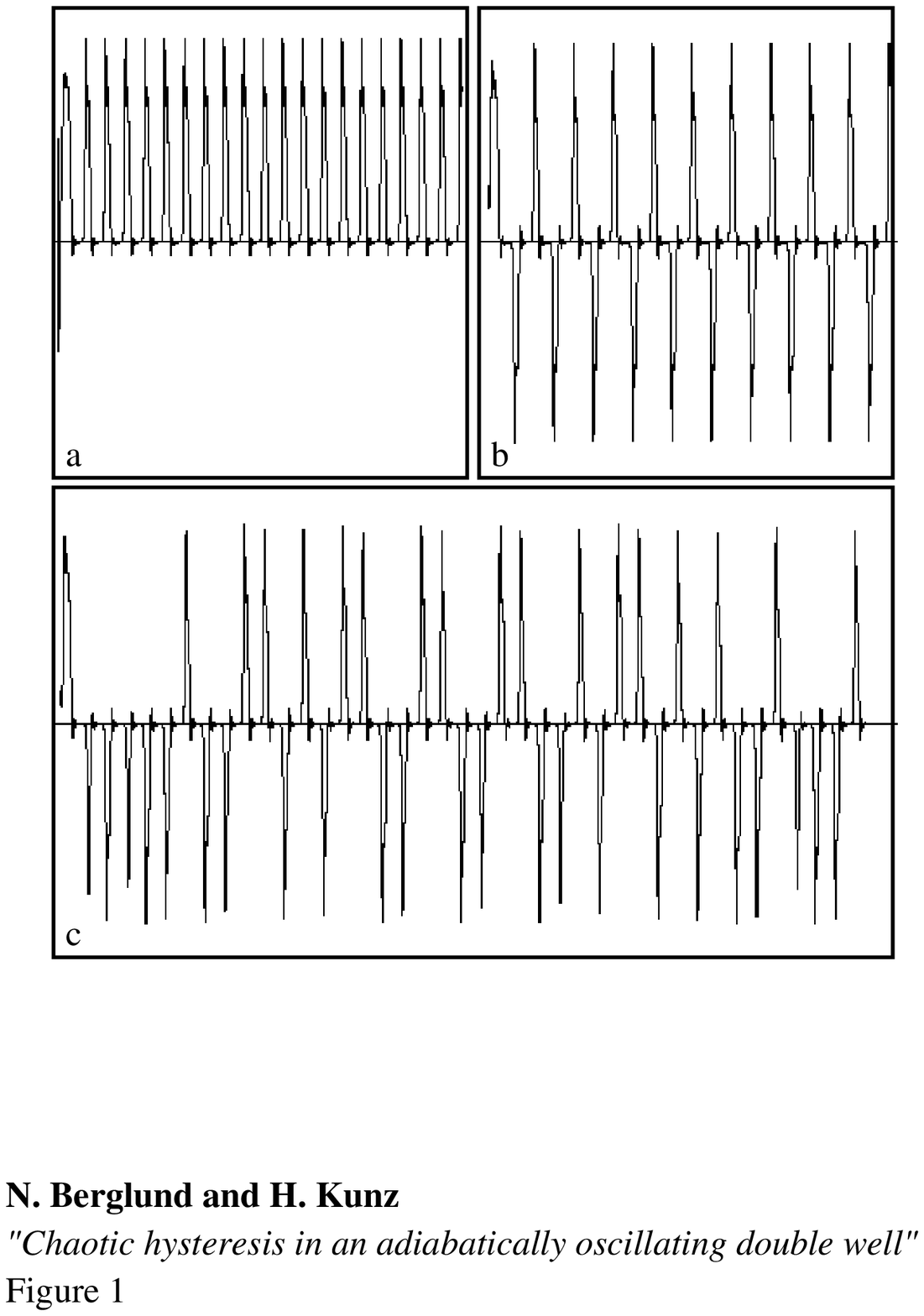,width=80mm,clip=t}}
\vspace{5mm}
\caption{Position of the particle (or the pendulum) as a function of time, 
	each peak corresponding to the particle falling into one of the 
	wells. The sequence of visited wells may be periodic, biperiodic 
	or chaotic.}
\end{figure}
	
The equation of motion of a damped particle in a slowly varying potential 
can be written
\begin{equation}
\ddot{q} + 2 \gamma \dot{q} + \Phi'(q,\lambda(\varepsilon t)) = 0,
\label{eq1}
\end{equation}
where the prime denotes derivation with respect to $q$ and 
$0<\varepsilon\ll 1$ is the adiabatic parameter.

Introducing the variables $x = (x_1,x_2) = (q,\dot{q})$, this equation can 
be transformed into the non--autonomous first order system
\begin{eqnarray}
\varepsilon\dot{x}_1 & = & x_2 \nonumber \\
\varepsilon\dot{x}_2 & = & - 2 \gamma x_2 - \Phi'(x_1,\lambda(\tau))
\label{eq2}
\end{eqnarray}
where the dots indicate now derivation with respect to the slow time 
$\tau=\varepsilon t$.

For fixed $\lambda(\tau)\equiv\lambda_0$, the dynamics of 
(\ref{eq2}) are well-known. Around the equilibria   
$x_1=q^*(\lambda_0)$, $x_2=0$, where $\Phi'(q^*(\lambda_0),\lambda_0)=0$, 
the linearization of (\ref{eq2}) has eigenvalues 
$a_\pm = -\gamma\pm\sqrt{\gamma^2-\phi''}$, with 
$\phi''=\Phi''(q^*(\lambda_0),\lambda_0)$. Hence the fixed point is a 
saddle if $\phi''<0$, a stable node if $0<\phi''<\gamma^2$ and a stable 
focus if $\phi''>\gamma^2$. With these informations, the phase portrait is 
easily drawn.

We will consider potentials of the following type: $\Phi(q,\lambda)$ is 
an even analytic function of $q$ such that the origin $O$ is hyperbolic 
when $\lambda>0$, a node for $\lambda_-(\gamma)<\lambda<0$ and a focus 
for $\lambda<\lambda_-$. For positive $\lambda$, $\Phi$ has two minima 
$\pm q^*(\lambda)$ which are nodes when $\lambda<\lambda_+(\gamma)$ 
and focuses when $\lambda>\lambda_+$.

The simplest example is the {\em Ginzburg--Landau 
potential} $\Phi(q,\lambda) = -\frac{1}{2}\lambda q^2+\frac{1}{4}q^4$ 
(here $\lambda_-=-\gamma^2$, $\lambda_+ = \gamma^2/2$), 
$q$ being the order parameter and $\lambda$ the difference between 
the temperature and its critical value.
But there is also a simple mechanical system which can be described by 
an equation of this type, namely a plane pendulum on a rotating table. 
In the frame rotating with frequency 
$\Omega$, it experiences a torque $-L M g \sin q$ due to its weight 
and a centrifugal torque $I\Omega^2\sin q\cos q$, where $L$ 
is the distance between suspension point $P$ and center of mass $G$, 
$I$ the moment of inertia with respect to $P$, and $q$ the angle 
between $PG$ and the vertical. Taking as a time unit 
$\Omega_{\mbox{$\scriptstyle\rm cr$}}^{-1} = (L M g/I)^{1/2}$, we obtain 
the equation of motion (\ref{eq1}) with 
$\Phi'(q,\lambda)=\sin q[1-(\lambda+1)\cos q]$, 
$\lambda = \Omega^2-1$.

Finally, the function $\lambda(\tau)$ we consider is periodic with period 1, 
and has exactly one minimum and one maximum. 
In order to analyze the behaviour of the dynamical system, we want to 
compute the Poincar\'e map in the $(x_1,x_2)$--plane during one period 
to dominant order in $\varepsilon$. This proceeds essentially by following 
the motion of $x$ along its static equilibria.
Let us denote by 
$a_{\pm}^o(\tau)$ and $a_{\pm}^*(\tau)$ the eigenvalues of the linearization 
around $O$ and $q^*(\lambda(\tau))$ respectively, and use the notation
\begin{eqnarray*}
\alpha^{o,*}(\tau_1,\tau_2) & = & 
\int_{\tau_1}^{\tau_2} \mbox{Re} \, a_+^{o,*}(\tau)\,d\tau,  \\
\phi^{o,*}(\tau_1,\tau_2) & = & 
\int_{\tau_1}^{\tau_2} \mbox{Im} \, a_+^{o,*}(\tau)\,d\tau,  \\
\delta^{o,*}(\tau_1,\tau_2) & = & 
\int_{\tau_1}^{\tau_2} \mbox{Re} (a_+^{o,*}-a_-^{o,*})(\tau)\,d\tau. 
\end{eqnarray*}

We first study (\ref{eq2}) in a neighborhood of the origin. The linearized 
system $\varepsilon\dot{x} = A(\tau)x$ can be solved by
constructing a linear change of variables 
$x = S(\tau,\varepsilon)y$ transforming the equation into 
$\varepsilon\dot{y} = B(\tau,\varepsilon)y$, where $B$ is as simple as 
possible. If $A(\tau)$ has distinct eigenvalues, one can find $S$ 
regular in a neighborhood of $\varepsilon=0$ such that $B$ has exponentially 
small off--diagonal terms. Complete diagonalization is possible if we 
only require $S$ to admit an asymptotic expansion in $\varepsilon$
\cite{W}. The full system 
(\ref{eq2}) thus becomes
\begin{eqnarray}
\varepsilon\dot{y}_1 & = & a^o_+(\tau,\varepsilon)y_1 
+ b_+(y_1,y_2;\tau,\varepsilon) 
\nonumber \\
\varepsilon\dot{y}_2 & = & a^o_-(\tau,\varepsilon)y_2 
+ b_-(y_1,y_2;\tau,\varepsilon),
\label{eq3}
\end{eqnarray}
with $a^o_\pm(\tau,\varepsilon) = a^o_\pm(\tau)+{\cal O}(\varepsilon)$ and 
$b_\pm = {\cal O}(|y|^3)$. We assume that $a^o_+(\tau_0)>0 $ 
and define an {\em adiabatic unstable manifold} 
$y_2=u(y_1;\tau,\varepsilon)$ by the equation
\[
\varepsilon\dot{u} = a^o_-u + b_-(y_1,u)-\partial_yu[a^0_+y_1+b_+(y_1,u)],
\]
where the initial condition coincides with the instantaneous unstable 
manifold at $\tau=\tau_0$. The solution, whose asymptotic 
expansion can be computed, may be continued 
to all times such that the origin is a saddle or a node. We then carry 
out the change of variables $y_2 = u(y_1)+\eta$ and define in a similar 
way an {\em adiabatic stable manifold} $y_1=v(\eta;\tau,\varepsilon)$. With 
$y_1 = v(\eta)+\xi$, (\ref{eq3}) becomes
\begin{eqnarray}
\varepsilon\dot{\xi} & = & [a^o_+(\tau,\varepsilon) 
+ \beta_+(\xi,\eta;\tau,\varepsilon)]\xi
 \nonumber \\
\varepsilon\dot{\eta} & = & [a^o_-(\tau,\varepsilon) 
+ \beta_-(\xi,\eta;\tau,\varepsilon)]\eta
\label{eq4}
\end{eqnarray}
with $\beta_\pm = {\cal O}(\xi^2+\eta^2)$.

\begin{figure}
\centerline{\psfig{figure=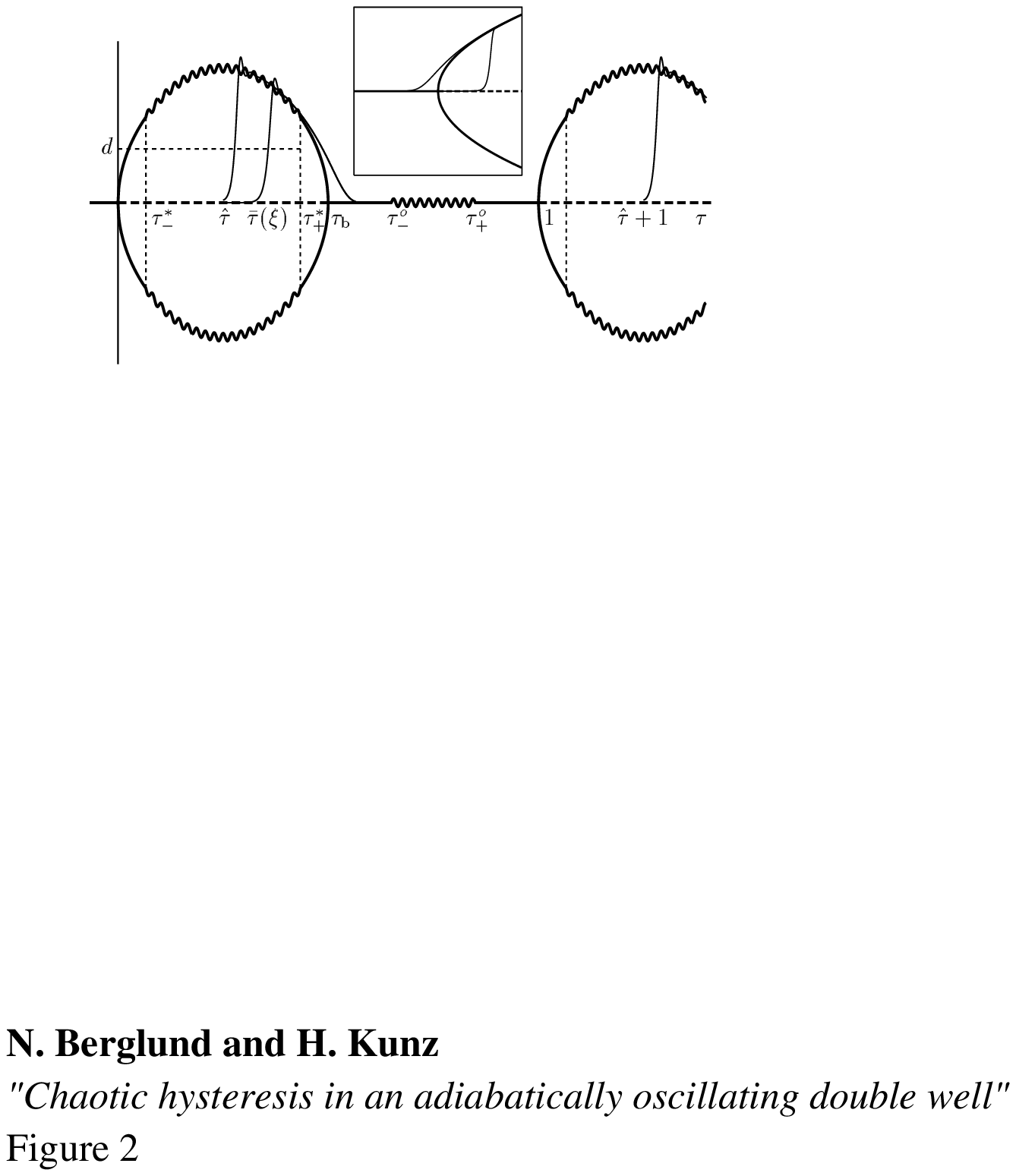,width=85mm,clip=t}}
\vspace{5mm}
\caption{Fixed points in the $(\tau,x_1)$--plane. Full lines represent nodes, 
	dashed lines hyperbolic points, and wavy lines focuses. The thin 
	lines are orbits of the system. The inset shows the static bifurcation 
	diagramm in the $(\lambda,x_1)$ plane (thick lines) with an 
	asymptotic hysteresis cycle (thin lines).}
\end{figure}
	
In the overdamped case, when $\lambda(\tau)$ always remains in 
the interval $(\lambda_-,\lambda_+)$, the problem is easy to analyze 
because it can be reduced 
to a one-dimensional one. Assume that $\lambda(\tau)>0$ for 
$0<\tau<\tau_{\mbox{$\scriptstyle\rm b$}}$ 
and $\lambda(\tau)<0$ for $\tau_{\mbox{$\scriptstyle\rm b$}}<\tau<1$. 
One shows that it is always 
possible to transform (\ref{eq2}) into (\ref{eq4}), with $a^o_-+\beta_-<0$.
Thus, we conclude from the second equation that $\eta(\tau)$ will go to 
zero exponentially fast, and it is sufficient to consider the reduced 
equation on the adiabatic manifold, 
$\varepsilon\dot{\xi}=
[a^o_++\beta_+(\xi,0)]\xi \equiv g(\xi;\tau,\varepsilon)$,
which undergoes a direct pitchfork bifurcation at 
$\tau={\cal O}(\varepsilon)$, and an inverse pitchfork bifurcation 
at $\tau=\tau_{\mbox{$\scriptstyle\rm b$}}+{\cal O}(\varepsilon)$.

The most important effect of the adiabaticity of the parameter 
variation is that the bifurcation is {\em delayed}. 
The orbit starting with a positive $\xi$ at 
$\tau_0\in(\tau_{\mbox{$\scriptstyle\rm b$}}-1,0)$ 
reaches the ${\cal O}(\varepsilon)$--neighborhood of the origin at 
$\tau_1=\tau_0+{\cal O}(\varepsilon|\ln\varepsilon|)$. As long as 
$\xi={\cal O}(\varepsilon)$, 
we have $\varepsilon\dot{\xi} = [a_+^o(\tau)+{\cal O}(\varepsilon)]\xi$ 
so that
\[
\xi(\tau) = \xi(\tau_1)\,\exp\frac{1}{\varepsilon}
[\alpha^o(\tau_1,\tau)+{\cal O}(\varepsilon)].
\]

The smallness of $\xi$ is thus guaranteed as long as $\alpha^o<0$.
For $\tau_1<\tau<0$, $\alpha^o$ is decreasing; it has increased again 
to zero only at $\tau=\Psi(\tau_1)>0$, which is by definition the 
{\em delayed bifurcation time}.
Afterwards, the orbit quickly jumps on the upper branch $q^*$ 
and follows it adiabatically until the inverse bifurcation at 
$\tau_{\mbox{$\scriptstyle\rm b$}}$,
where we can prove that 
$\xi(\tau_{\mbox{$\scriptstyle\rm b$}})=C\varepsilon^{1/4}$. 
The next delayed bifurcation occurs at 
$\hat{\tau}=\Psi(\tau_{\mbox{$\scriptstyle\rm b$}})$.

It follows that the Poincar\'e map, defined by 
$\xi(\hat{\tau}+1)=T(\xi(\hat{\tau}))$, 
is a monotonic odd function of $\xi$, such that for 
$\xi>\exp-\frac{1}{\varepsilon}\alpha^o(\hat{\tau},\hat{\tau}+1)$, one has 
$0<c_1\varepsilon^{1/4}<T(\xi)<c_2\varepsilon^{1/4}$. 
This implies that the map has three fixed points, the unstable origin 
and two symmetric stable points. The positive solution corresponds to 
the attractive periodic orbit of Fig.\ 1a, and, if plotted in the 
$(\lambda,\xi)$--plane, to an asymptotic hysteresis cycle (see inset 
of Fig.\ 2), whose area we are able to prove scales like  
${\cal A}(\varepsilon)\simeq{\cal A}(0)+c\varepsilon^{3/4}$.

We now turn to the most complicated case, when the amplitude of 
$\lambda(\tau)$ is large enough for the origin to be a focus 
for $\tau_-^o<\tau<\tau_+^o$, and for $q^*$ to be a focus for 
$\tau_-^*<\tau<\tau_+^*$. We make a Poincar\'e section at the delay time 
$\hat{\tau}=\Psi(\tau_{\mbox{$\scriptstyle\rm b$}}-1)$, which we assume 
to be in the interval $(\tau_-^*,\tau_+^*)$ (Fig.\ 2).

The variables $\zeta=(\xi,\eta)$ are defined for 
$\tau\notin[\tau^o_-,\tau^o_+]$ and for $|\xi|<d$ where $d$ is a constant 
such that $x_1$ is closer to the origin than the focus at $q^*$. 
We want to compute $\zeta(\hat{\tau}+1)=(\xi_1,\eta_1)$ as a function of 
$\zeta(\hat{\tau}) = (\xi_0,\eta_0)$, in the case where 
$\xi_0\in(0,d)$ and $\eta_0={\cal O}(\varepsilon)$. 
From (\ref{eq4}), we deduce that  $\xi(\tau)=d$ 
at $\tau=\bar{\tau}(\xi_0)+{\cal O}(\varepsilon)$, where $\bar{\tau}(\xi_0)$ 
is the solution of 
\begin{equation}
\alpha^o(\hat{\tau},\bar{\tau}) = -\varepsilon\ln(\xi_0/d).
\label{taubar}
\end{equation}

For most trajectories, $\bar{\tau}$ is very close to $\hat{\tau}$, 
but orbits which are exponentially close to the origin at $\hat{\tau}$ 
can be appreciably delayed. $\bar{\tau}(\xi_0)$ is a decreasing function 
of $\xi_0$ for $\xi_{\mbox{$\scriptstyle\rm c$}}\leq\xi_0\leq d$, with 
$\bar{\tau}(d)=\hat{\tau}$ 
and $\bar{\tau}(\xi_{\mbox{$\scriptstyle\rm c$}}) = \tau^*_+$, where 
$\xi_{\mbox{$\scriptstyle\rm c$}}=
d\exp-\frac{1}{\varepsilon}\alpha^o(\hat{\tau},\tau^*_+)$.

Next, we construct a particular solution of (\ref{eq2}) $\bar{x}(\tau)$, 
wich remains in a neighborhood of the upper branch $q^*(\tau)$ and 
such that $\bar{\xi}(\tau_{\mbox{$\scriptstyle\rm b$}})
=C\varepsilon^{1/4}$, $\bar{\eta}(\tau_{\mbox{$\scriptstyle\rm b$}})=0$. 
Writing $x=\bar{x}(\tau) + y$, we obtain the equation 
$\varepsilon\dot{y}=\bar{A}(\tau;\varepsilon)y+b(y,\tau)$, where $\bar{A}$ 
has eigenvalues $a_{\pm}^*(\tau)+{\cal O}(\varepsilon^{1/2})$. The nonlinear 
term $b(y,\tau)$ may be decreased to order $\varepsilon$ by the change 
of variables $y=z+\chi(z)$, which puts the instantaneous system 
into normal form. When 
$\tau\neq\tau^*_+$, the linear part of the equation can be 
diagonalized. Then we have to take care of the turning point $\tau^*_+$, 
where the eigenvalues of $\bar{A}(\tau)$ cross and $S(\tau)$ diverges.
The problem is solved by carrying out, in the interval 
$[\tau^*_+-\varepsilon^{2/3},\tau^*_++\varepsilon^{2/3}]$, a change of 
variables which puts $B$ into Jordan form, so that one obtains 
Airy's equation. 
Putting together these steps, we finally obtain 
\begin{eqnarray}
\xi(\tau_{\mbox{$\scriptstyle\rm b$}}) & = & 
\bar{\xi}(\tau_{\mbox{$\scriptstyle\rm b$}}) + 
\mbox{e}^{\alpha_*/\varepsilon}\sin(\phi_*/\varepsilon)
 \nonumber \\
\eta(\tau_{\mbox{$\scriptstyle\rm b$}}) & = & 
\mbox{e}^{(\alpha_*-\delta^*)/\varepsilon}\sin(\phi_*/\varepsilon+\theta^*), 
\label{sols}
\end{eqnarray}
where $\delta^*=\delta^*(\tau^*_+,1)+{\cal O}(\varepsilon^{1/2})$ is a 
constant and $\alpha_*=\alpha_*(\xi_0)+{\cal O}(\varepsilon^{1/2})$, 
$\phi_*=\phi_*(\xi_0)+{\cal O}(\varepsilon)$, with
\begin{eqnarray}
\alpha_*(\xi_0) & = & \alpha^*(\bar{\tau}(\xi_0),1) \nonumber \\
\phi_*(\xi_0) & = & \phi^*(\bar{\tau}(\xi_0),\tau^*_+). 
\label{apstar}
\end{eqnarray}
Of course $\zeta(\tau_{\mbox{$\scriptstyle\rm b$}})$ depends also on 
$\eta_0$ via $y(\bar{\tau})$, 
but only at next--to--leading order. For 
$\xi_0\in[\xi_{\mbox{$\scriptstyle\rm c$}},d]$, (\ref{sols}) 
is the parametric equation of an exponentially squeezed spiral.

Proceeding in a similar way around the origin, we find  
$\zeta(\hat{\tau}+1) = U\zeta(\tau_{\mbox{$\scriptstyle\rm b$}})$, with
\begin{equation}
U = 
\left(
\begin{array}{lr}
\phantom{\mbox{e}^{-\delta^o_2/\varepsilon}}  
\sin\left(\frac{\phi^o}{\varepsilon}\right) & 
\mbox{e}^{-\delta^o_2/\varepsilon} 
\sin\left(\frac{\phi^o}{\varepsilon}+\theta^o_2\right)\\
& \\
\mbox{e}^{-\delta^o_1/\varepsilon} 
\sin\left(\frac{\phi^o}{\varepsilon}+\theta^o_1\right) & 
\mbox{e}^{-\delta^o_3/\varepsilon} 
\sin\left(\frac{\phi^o}{\varepsilon}+\theta^o_3\right)
\end{array}
\right),
\label{sol0}
\end{equation}
where $\phi^o = \phi^o(\tau^o_-,\tau^o_+) + {\cal O}(\varepsilon^{1/2})$ 
is the dynamic phase, 
$\delta^o_1 = \delta^o(0,\tau^o_-) + {\cal O}(\varepsilon^{1/2})$,
$\delta^o_2 = \delta^o(\tau^o_+,\hat{\tau}) + {\cal O}(\varepsilon^{1/2})$,
and $\delta^o_3 = \delta^o_1 + \delta^o_2$.
The geometric phase shifts  
$\theta_i$ can be expressed in terms of Airy functions (plus 
corrections of order $\varepsilon^{1/3}$). This transformation acts like 
a rotation of angle $\phi^o/\varepsilon$ and an exponential contraction along 
the stable manifold.

Combining (\ref{sols}) and (\ref{sol0}), we 
finally obtain the expression of the Poincar\'e map 
$\xi_1 = T_1(\xi_0,\eta_0;\varepsilon)$, 
$\eta_1 = T_2(\xi_0,\eta_0;\varepsilon)$
with
\begin{eqnarray}
T_1 & = & \sin\left(\frac{\phi^o}{\varepsilon}\right)
\left[C\varepsilon^{1/4} + \mbox{e}^{\alpha_*/\varepsilon}
\sin\left(\frac{\phi_*}{\varepsilon}\right)\right] \nonumber \\
& + & \mbox{e}^{(\alpha_*-\delta^*-\delta^o_2)/\varepsilon}
\sin\left(\frac{\phi^o}{\varepsilon}+\theta^o\right)
\sin\left(\frac{\phi_*}{\varepsilon}+\theta^*\right)
\label{poincare}
\end{eqnarray}
and $T_2 = {\cal O}(\mbox{e}^{-\delta^o_1/\varepsilon})$. 

The expression (\ref{poincare}) of $T_1$ is valid for 
$\xi_0>\xi_{\mbox{$\scriptstyle\rm c$}}$. 
By symmetry, $T_1$ is an odd function of $\xi_0$, and for 
$|\xi_0|<\xi_{\mbox{$\scriptstyle\rm c$}}$, 
it is monotonic as in the overdamped case. 
All variables appearing in (\ref{poincare}) are independent of 
$\xi_0,\,\eta_0$ at lowest order in $\varepsilon$, except $\alpha_*$ 
and $\phi_*$ which are given by (\ref{apstar}).

Since $\eta$ is exponentially contracted at each iteration, we may 
replace the two--dimensional invertible Poincar\'e map by the 
non--invertible one--dimensional map $\xi_0\mapsto T_1(\xi_0,0)$. 
Its graph is oscillating  around 
$C\varepsilon^{1/4}\sin(\phi^o/\varepsilon)$ 
with increasing amplitude and frequency as 
$\xi\searrow\xi_{\mbox{$\scriptstyle\rm c$}}$. 
One can prove that there is a positive constant $\mu$ such that 
when $\sin(\phi^o/\varepsilon)>\mbox{e}^{-\mu/\varepsilon}$, 
$T_1$ admits only one positive fixed point 
at $\xi\sim\varepsilon^{1/4}\sin(\phi^o/\varepsilon)$, which is stable. 
In this case, we obtain a 
hysteresis cycle of period 1. Similarly, when 
$\sin(\phi^o/\varepsilon)<-\mbox{e}^{-\mu/\varepsilon}$, 
the map has a stable orbit 
of period 2, corresponding to the particle falling alternatively 
into the left and right well.

The most interesting situation occurs when 
$|\sin(\phi^o/\varepsilon)|$ $<$ $\mbox{e}^{-\mu/\varepsilon}$. In this case, 
we observed numerically a great variety of behaviours, including period 
doubling cascades and chaotic orbits. These ``chaotic'' zones occur at 
integer values of $k=\phi^o/\pi\varepsilon$, and the width and height of 
the $n$th zone are proportional to $\mbox{e}^{-(\mu\pi/\phi^o)n}$. 
This prediction has been confirmed numerically.

The map (\ref{poincare}) certainly deserves further study. But the most 
important fact to us is that it explains accurately the alternance 
of periodic, biperiodic and chaotic hysteresis, in accordance with 
numerical simulations (Fig.\ 3) as well as with the laboratory experiment 
of the rotating pendulum.

We thank P. Braissant and B. Egger for carrying 
out the laboratory experiment of the rotating pendulum. This work is 
supported by the Fonds National Suisse de la Recherche Scientifique.

\begin{figure}
\centerline{\psfig{figure=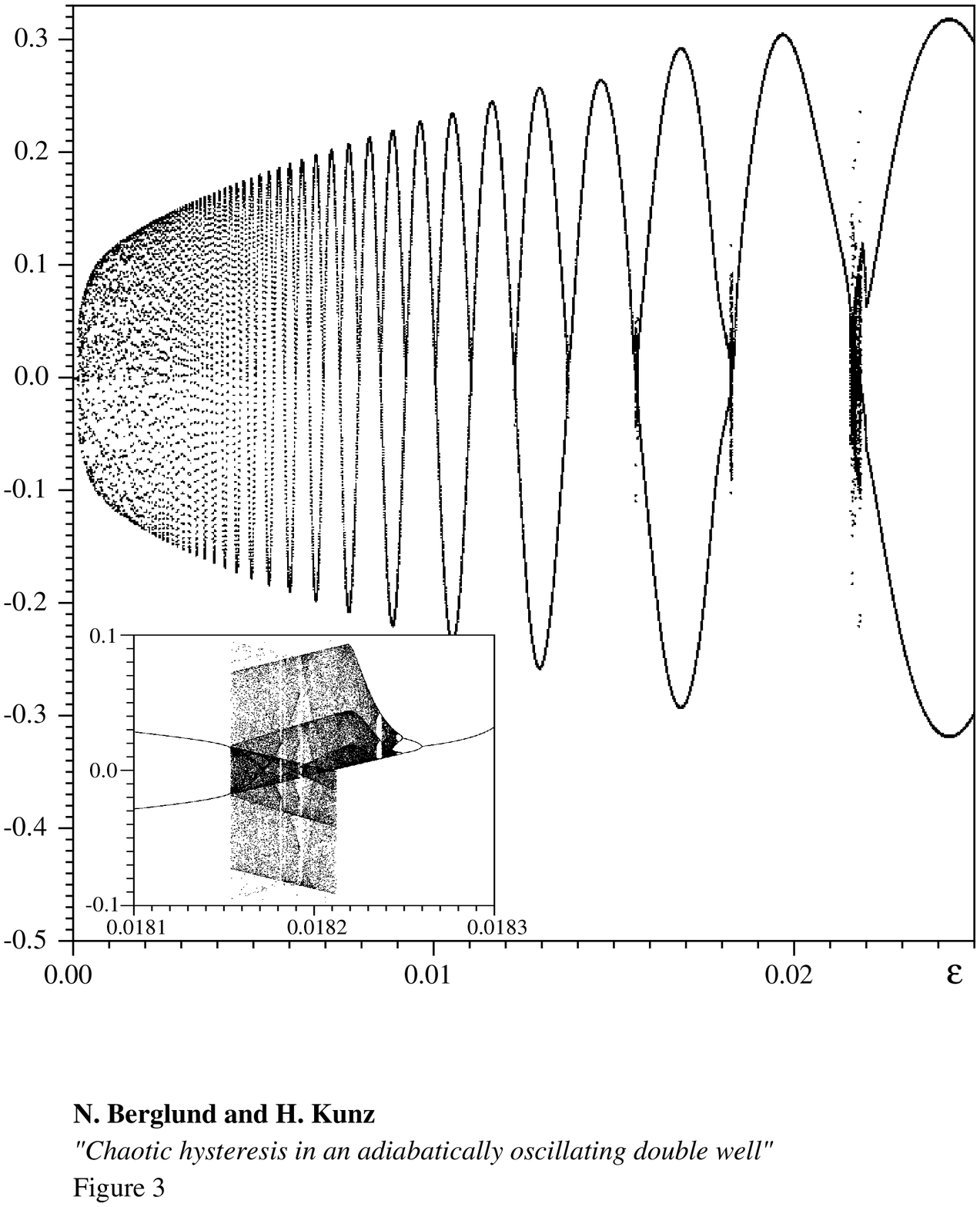,width=85mm,clip=t}}
\vspace{5mm}
\caption{Numerically computed bifurcation diagramm of the Poincar\'e map 
	for the rotating pendulum. For each value of $\varepsilon$ on the 
	abscissa, we plot the asymptotic behaviour of $\xi(\hat{\tau}+n)$, 
	$n\in\mbox{I}\!\mbox{N}$, for {\em one} initial condition. 
	Regions with a period--1 and period--2 cycle 
	are separated by small chaotic zones, the inset showing an 
	enlargement of the second zone from the right.} 
\end{figure}

\end{document}